\documentclass[twocolumn]{aastex63}
\usepackage{placeins}

\received{November 23, 2021}
\revised{June 13, 2022}
\accepted{June 24, 2022}

\submitjournal{ApJ}

\shorttitle{Phosphorus Molecules Towards Low-Mass Protostars}
\shortauthors{Wurmser \& Bergner}

\graphicspath{{./}{figures/}}

\begin{document}

\title{New Detections of Phosphorus Molecules towards Solar-type Protostars}

\author[0000-0003-4791-5331]{Serena Wurmser}
\affiliation{Center for Astrophysics $|$  Harvard \& Smithsonian, 60 Garden St., Cambridge, MA 01238, USA}

\author[0000-0002-8716-0482]{Jennifer B. Bergner}
\altaffiliation{NASA Sagan Fellow}
\affiliation{University of Chicago Department of the Geophysical Sciences, Chicago, IL 60637, USA}

\begin{abstract}
Phosphorus is a necessary element for life on Earth, but at present we have limited constraints on its chemistry in star- and planet-forming regions: to date, phosphorus carriers have only been detected towards a few low-mass protostars.  Motivated by an apparent association between phosphorus molecule emission and outflow shocking, we used the IRAM 30m telescope to target PN and PO lines towards seven Solar-type protostars with well-characterized outflows, and firmly detected phosphorus molecules in three new sources.  This sample, combined with archival observations of three additional sources, enables the first exploration of the demographics of phosphorus chemistry in low-mass protostars.  The sources with PN detections show evidence for strong outflow shocks based on their H$_2$O 1$_{10}$--1$_{01}$ fluxes.  On the other hand, no protostellar properties or bulk outflow mechanical properties are found to correlate with the detection of PN.  This implies that gas-phase phosphorus is specifically linked to shocked gas within the outflows.  Still, the PN and PO line kinematics suggest an emission origin in post-shocked gas rather than directly shocked material.  Despite sampling a wide range of protostellar properties and outflow characteristics, we find a fairly narrow range of source-averaged PO/PN ratios (0.6--2.2) and volatile P abundances as traced by (PN+PO)/CH$_3$OH ($\sim$1-3\%).  Spatially resolved observations are needed to further constrain the emission origins and environmental drivers of the phosphorus chemistry in these sources.
\end{abstract}


\section{Introduction} \label{sec:intro}
Along with H, C, O, N, and S, phosphorus is considered to be a key biogenic element and is found in all known forms of life on Earth. It is a critical component in DNA and RNA, ATP, and a host of other biological molecules \citep{Macia2005}. While the biological importance of this element is well established, there remain many questions about how phosphorus was distributed during the formation of our Solar System. Given its importance for terrestrial life, phosphorus may also play a central role in the development of life on other planets. Ultimately, understanding the speciation and distribution of phosphorus in other nascent planetary systems can help us to contextualize Solar System constraints on the phosphorus chemistry, and to explore prospects for prebiotic chemistry in systems other than our own.

The first phosphorus molecule detected in a star forming region was phosphorus nitride (PN). \cite{Turner1987} presented PN detections in Ori (KL), W51M, and Sgr B2---three high-mass star-forming regions (SFRs)---and proposed that the higher-than-predicted gas-phase abundances were the result of grain disruption. While phosphorus molecules were subsequently detected around evolved stars \citep[e.g.][]{Guelin1990, Agundez2007, Tenenbaum2007}, it was not until 2011 when \citeauthor{Yamaguchi2011} presented the first detection of PN in a low-mass (i.e.~solar-type) SFR, L1157. This was soon followed by the first detection of PO in a low-mass SFR, towards the same source L1157 \citep{Lefloch2016}.  In recent years, PN and PO have been detected in a host of other massive SFRs and molecular clouds \citep[e.g.][]{Rivilla16,Mininni2018, Rivilla2018, rivilla20, Bernal21}.  In low-mass SFRs, PN detections were reported towards three sources as part of the TIMASS and ASAI surveys \citep{Caux2011, Lefloch2018}, and PN and PO were both detected towards the B1-a protostar \citep{bergner19}.  Characterizing the phosphorus molecule emission in low-mass star forming regions is of particular importance because these systems are analogs to the proto-Solar Nebula.

While the chemistry of PO and PN in SFRs remains poorly understood, several common themes have emerged. In all low- and high-mass SFRs where PO and PN have both been detected, the source-averaged PO/PN ratio is within the range of 1--3 \citep{Lefloch2016, Rivilla16,Rivilla2018,bergner19, Bernal21}. Correlations between the line profiles and column densities of PN and PO with the shock tracer SiO support that gas-phase P molecules are typically a product of grain sputtering in shocked regions \citep{Lefloch2016, Mininni2018, Rivilla2018, fontani2019, bergner19, Bernal21}. While phosphorus chemical models predict that PO and PN both form in the gas phase following the desorption of PH$_3$ from grains, no other P molecules including PH$_3$ have been detected in SFRs with PO and PN detections \citep{Yamaguchi2011, Lefloch2016, Jimenez-Serra2018}.

In the Solar System, the phosphorus abundance in primitive meteorites is close to Solar, indicating a large reservoir of refractory phosphorus in the Solar Nebula  \citep[most commonly in the form of phosphate minerals;][]{Lodders2003}.  Additionally, the volatile phosphorus carrier PO was recently detected towards comet 67P/Churyumov–Gerasimenko (hereafter 67P) by the Rosetta mission \citep{Altwegg16, rivilla20} with a high elemental P/O ratio $\sim$0.1--0.5$\times$ the Solar value.  This is especially noteworthy given that comets may have played a role in the delivery of prebiotic material to young terrestrial planets \citep[e.g.][]{Rubin2019a}.  The PO/PN ratio in comet 67P was found to be at least 10.  Whether the Solar System is typical in its volatile P abundance and composition remains an open question.

The small sample size of low-mass SFRs with P molecule detections is a major obstacle to constraining the phosphorus astrochemistry relevant to planet formation.  Informed by growing evidence that outflow shocks play a key role in releasing phosphorus in the gas in low-mass star-forming regions \citep{Yamaguchi2011, Lefloch2016, bergner19, bergner21}, we targeted PO and PN lines towards seven low-mass protostars with well-characterized outflows using the IRAM 30m telescope.   In this paper, we present detections of phosphorus carriers towards three new low-mass protostars (NGC1333-IRAS 3, Ser-SMM1, Ser-SMM4), with one additional tentative detection (L723). This sample is combined with archival observations of B1-b and NGC1333-IRAS 4a from the ASAI program \citep{Lefloch2018}, for which PN detections were previously reported but not characterized.  Section \ref{sec:observations} describes our observations, including the source and line targets.  In Section \ref{sec:methods} we outline our analysis methods, and Section \ref{sec:results} presents the resulting PO and PN column densities and column density ratios with respect to CH$_3$OH and H$_2$. With our expanded sample of phosphorus molecules in Solar-type star forming regions, in Section \ref{sec:discussion} we discuss the trends identified across our sample, and comparisons between the P inventories in low-mass protostars and primitive Solar System materials.

\section{Observations} \label{sec:observations}
\subsection{Source targets} \label{subsec:sources}
Our source sample (Table \ref{tab:targets}) is selected from the embedded low-mass protostars observed by the Herschel WISH program.  For these sources, the outflow properties have been previously characterized based on the H$_2$O ground-state 1$_{10}$--1$_{01}$ transition and auxiliary CO observations \citep{kristensen12, Yildiz2015}.  Of the sources observable by the IRAM 30m telescope, our targets were chosen to span a range of physical properties (bolometric luminosities from $\sim$2--40 L$_\odot$, envelope masses from $\sim$1--50 M$_\odot$) as well as evolutionary stages (bolometric temperatures from $\sim$20--150 K).  

\begin{deluxetable*}{lcccccccc}
\tablenum{1}
\tablecaption{Source Targets\label{tab:targets}}
\tablewidth{0pt}
\tablehead{ \colhead{Source} & \colhead{RA} & \colhead{Dec} & \colhead{Dist.} & \colhead{v$_\mathrm{lsr}$} &  \colhead{T$_\mathrm{bol}$} & \colhead{L$_\mathrm{bol}$} & \colhead{M$_\mathrm{env}$} & \colhead{Class} 
\\
\nocolhead{} & \colhead{(h m s)} & \colhead{($^\circ$, ', ")} & \colhead{(pc)}& \colhead{(km s$^{-1}$)}& \colhead{(K)} & \colhead{(L$_\odot$) } & \colhead{(M$_\odot$) } & \colhead{}
}
\startdata
Ser-SMM4$^*$	&	18 29 56.6	&	+01 13 15.1	&	415 & 7.8	&	26	&	6.2	&		6.9	&	0	\\
Ser-SMM3$^*$	&	18 29 59.2	&	+01 14 00.3	&	415 & 7.6 &	38	&	16.6	&	10.4	&	0	\\
Ser-SMM1$^*$	&	18 29 49.8	&	+01 15 20.5	&	415 & 8.5	&	39	&	99	&	52.5	&	0	\\
L723$^*$	&	19 17 53.7	&	+19 12 20.0	&	300 & 11	&	    39	&	3.6	&	1.3	&	0		\\
L1527$^*$	&	04 39 53.9	&	+26 03 09.8	&	140 & 5.9	&	    44	&	1.9	&	0.9	&	0		\\
L1551-IRS5$^*$	&	04 31 34.1	&	+18 08 05.0	&	140 & 6.5	&	94	&	22.1	&	2.3	&	I		\\
NGC1333-IRAS 3$^*$	&	03 29 03.8	&	+31 16 04.0	&235 & 8.3 &	149	&	41.8	&	8.6	&	I		\\
\hline
B1-b	&	03 33 21.0	&	+31 07 30.7	&	300	& 6.8 &	16	&	0.7	&	26.1	&	0		\\
NGC1333-IRAS 4A$^*$	&	03 29 10.5	&	+31 13 30.9	&	235 & 7.2	&	33	&	9.1	&	5.2	&	0		\\
B1-a	&	03 33 16.7	& +31 07 55.1	&	300 &	6.8 &	158	&	1.3	&	2.8	&	I		\\
\enddata
\tablecomments{Properties for the WISH sources (noted with $^*$) are taken from \citet{Mottram2014}.  Properties for B1-b and B1-a are taken from  \cite{Pezzuto2012} and \cite{Hatchell2007}. Because B1-b is a binary system, we list the average T$_\mathrm{bol}$ and total L$_\mathrm{bol}$ of the two components.  Sources with new coverage of phosphorus molecule lines are listed above the horizontal line, and those with archival coverage are below the line.}
\end{deluxetable*}

Importantly, the sources also show different outflow characteristics.  Mechanical properties of the outflows derived from CO 3-2 emission \citep{Yildiz2015} span a range of values, with outflow masses from $\sim$8$\times$10$^{-3}$--1.5$\times$10$^{-1}$ M$_\odot$, outflow forces from $\sim$2$\times$10$^{-4}$--2$\times$10$^{-3}$ M$_\odot$ yr$^{-1}$ km s$^{-1}$, and dynamical ages from 1--15 kyr.  The water emission \citep{kristensen12}, which traces shocked gas, spans a range of line widths ($\sim$20--80 km s$^{-1}$) and velocity-integrated intensities (1--18 K km s$^{-1}$).  Also, the water lines in the sample exhibit different kinematic components: broad (FWHM$>$20 km s$^{-1}$), medium (FWHM=5--20 km s$^{-1}$), or both broad and medium. Note that all sources additionally exhibit a narrow (FWHM$<$5 km s$^{-1}$) emission/absorption component.  With this sample diversity, we aim to identify what protostellar and/or outflow properties are associated with the emission of gas-phase P molecules.

We note that the Herschel H$_2$O 1$_{10}$--1$_{01}$ observations correspond to a beam size of 39\arcsec, which is larger than the IRAM 30m beam FWHM (16--27\arcsec at 150-90 GHz).  Higher-frequency water lines (corresponding to smaller beam sizes of $\sim$13--28\arcsec) were also detected by Herschel towards the WISH sources \citep{Mottram2014}, confirming that shocks are present on the smaller spatial scales accessed by the IRAM 30m beam.

\subsection{Description of observations}
Observations of our source targets were taken with the IRAM 30m telescope using the EMIR 90 GHz (3mm) and 150 GHz (2mm) receivers with the Fourier Transform Spectrometer backend.  The 3mm setting was observed with a spectral resolution of 200 kHz, and the 2mm setting was observed with a resolution of 50 kHz.  Observations were taken between 2020 March 03 and 2020 March 06.  Additional details on the observing parameters and quality can be found in Appendix \ref{app:obsdeets}.  Initial data reduction steps were performed using CLASS \footnote{https://www.iram.fr/IRAMFR/GILDAS/}, assuming a main-beam efficiency described by the Ruze equation with a surface RMS of 66$\mu$m and B$_\mathrm{eff,0}$=0.863, as appropriate for the IRAM 30m telescope.  Spectral line fitting and subsequent analysis were performed with Python.  

\subsection{Line Targets}
Every source was observed in the 2mm setup, the 3mm setup, or both the 2mm and 3mm setups, as a result of variable weather conditions (Table \ref{tab:observations}).  Each setup covers one PN line and four PO hyperfine components.  In addition, C$^{17}$O 1--0 was covered in the 3mm setup, and a range of CH$_3$OH and SO lines were covered in both the 2mm and 3mm setups. The line parameters used in our analysis are summarized in Table \ref{tab:line_targets}.  Additional details on the spectral setups can be found in Appendix \ref{app:obsdeets}.

\subsection{Additional sources}
In addition to the new observations described above, we also made use of archival IRAM 30m observations of NGC1333-IRAS 4a and B1-b from the ASAI program \citep{Lefloch2018}.  PN was previously reported as detected towards these sources, however there has not yet been an analysis of the PN (or PO) column densities or line kinematics.  ASAI provided coverage of the 3mm and 2mm PN and PO lines targeted by our survey, as well as higher-frequency (1.3mm) lines.  The observational details can be found in \citet{Lefloch2018}.

We also include the previous constraints on PN and PO emission in B1-a from \citet{bergner19} when comparing across the source sample (Section \ref{subsec:sample_trends}).  Note that B1-a and B1-b were not part of the WISH program.

\begin{deluxetable}{l|rccc}
\tablenum{2}
\tablecaption{Line Targets\label{tab:line_targets}}
\tablewidth{0pt}
\tablehead{ \colhead{Molecule} & \colhead{Frequency (GHz)} & \colhead{E$_u$ (K)} &  \colhead{g$_u$} & \colhead{Log$_{10}$(A$_{ul}$)}
}
\startdata
PN	&	93.9798901	&	6.8	&	5	&  -4.0\\
    &	140.9676931	&	13.5	&	7  &  -4.5\\
    &   $^+$234.9356952 &   33.8    &   11  & -3.2 \\
\hline
PO	 &	$^f$*107.206200	&	8.4	&	7  &    -4.7\\
    &	108.998445	&	8.4	&	7  & 	-4.7\\
	&	109.045396	&	8.4	&	5  & 	-4.7\\
	&	$^f$109.281189	&	8.4	&	5	&   -4.7\\
    &	$^f$*151.855454	&	15.8	&	9 &   -4.2	\\
	&	152.656979	&	15.7	&	9 &  -4.2\\
    &	152.680282	&	15.7	&	7 &  -4.2\\
    &	$^f$152.888128	&	15.7	&	7  &  -4.7	\\
    &   $^+$239.948982 & 36.7 & 13   & -3.6  \\
    &   $^+$239.958101 & 36.7 &  11 & -3.6 \\
    &   $^{f+}$*240.141059 & 36.7 & 13   & -3.6 \\
    &   $^{f+}$240.152528 & 36.7 & 11   & -3.6 \\
\hline
E-CH$_3$OH	&	95.169391	&	83.5	&	68	&	-5.4	\\
	&	95.914310	&	21.4	&	20	&	-5.6	\\
	&	96.755501	&	28.0	&	20	&	-5.6	\\
	&	108.893945	&	13.1	&	4	&	-4.8	\\
	&	156.127544	&	86.5	&	52	&	-5.2	\\
	&	156.488902	&	96.6	&	68	&	-4.7	\\
	&	156.602395	&	21.4	&	20	&	-4.7	\\
	&	156.828517	&	78.1	&	60	&	-4.7	\\
	&	157.048617	&	61.8	&	52	&	-4.7	\\
	&	157.178987	&	47.9	&	44	&	-4.7	\\
	&	157.246062	&	36.3	&	36	&	-4.7	\\
	&	157.276019	&	20.1	&	20	&	-4.7	\\
\hline
SO	&	99.29987	&	9.2 &	7	&	-4.9	\\
	&	138.1786	&	15.9	&	9	&	-4.4	\\
\hline
C$^{17}$O & 112.358777 & 5.39 & 4 & -7.17 \\
          & 112.358982 & 5.39 & 8 & -7.17 \\
          & 112.360007 & 5.39 & 6 & -7.17 \\
	\enddata
\tablenotetext{}{Line parameters are from CDMS \citep{Muller2001, Muller2005} based on data from \citet{Cazzoli2006}, \citet{Bailleux2002}, \citet{Kawaguchi1983}, \citet{Xu2008}, \citet{Clark1976}, and \citet{Klapper2003}. $^+$ indicates lines that are covered only for the ASAI sources.  Note that we include only E-CH$_3$OH lines in our analysis.  PO hyperfine lines of a given J-level were stacked after normalizing to those marked with a *.  $^f$ indicates an $f$-type PO transition, while all other PO lines are $e$.}
\end{deluxetable}

\section{Column Density Constraints} \label{sec:methods}

\subsection{Line Fitting and Detections}
\label{sec:linefits}
For each line target, we subtracted a local baseline and fit a Gaussian to the spectrum.  In the case of PO, we stacked the four hyperfine components of each rotational level to increase the SNR. This was done by producing spectra centered on each hyperfine line and rebinned to a common velocity grid.  Each hyperfine spectrum was scaled based on its intrinsic line strength relative to the strongest hyperfine component (marked with $^*$ in Table \ref{tab:line_targets}), then all spectra were averaged.  The PN and PO Gaussian fit results, including velocity-integrated intensities, are listed in Table \ref{tab:p_gaussian_fitting}, and those for CH$_3$OH are in Table \ref{tab:CH3OH_detections}.  Uncertainties on the integrated intensities consist of a 10\% calibration uncertainty added in quadrature with the Gaussian fit uncertainty, and are propagated through subsequent analysis steps.

\begin{figure*}[t]
\begin{center}
	{\includegraphics[width=\textwidth]{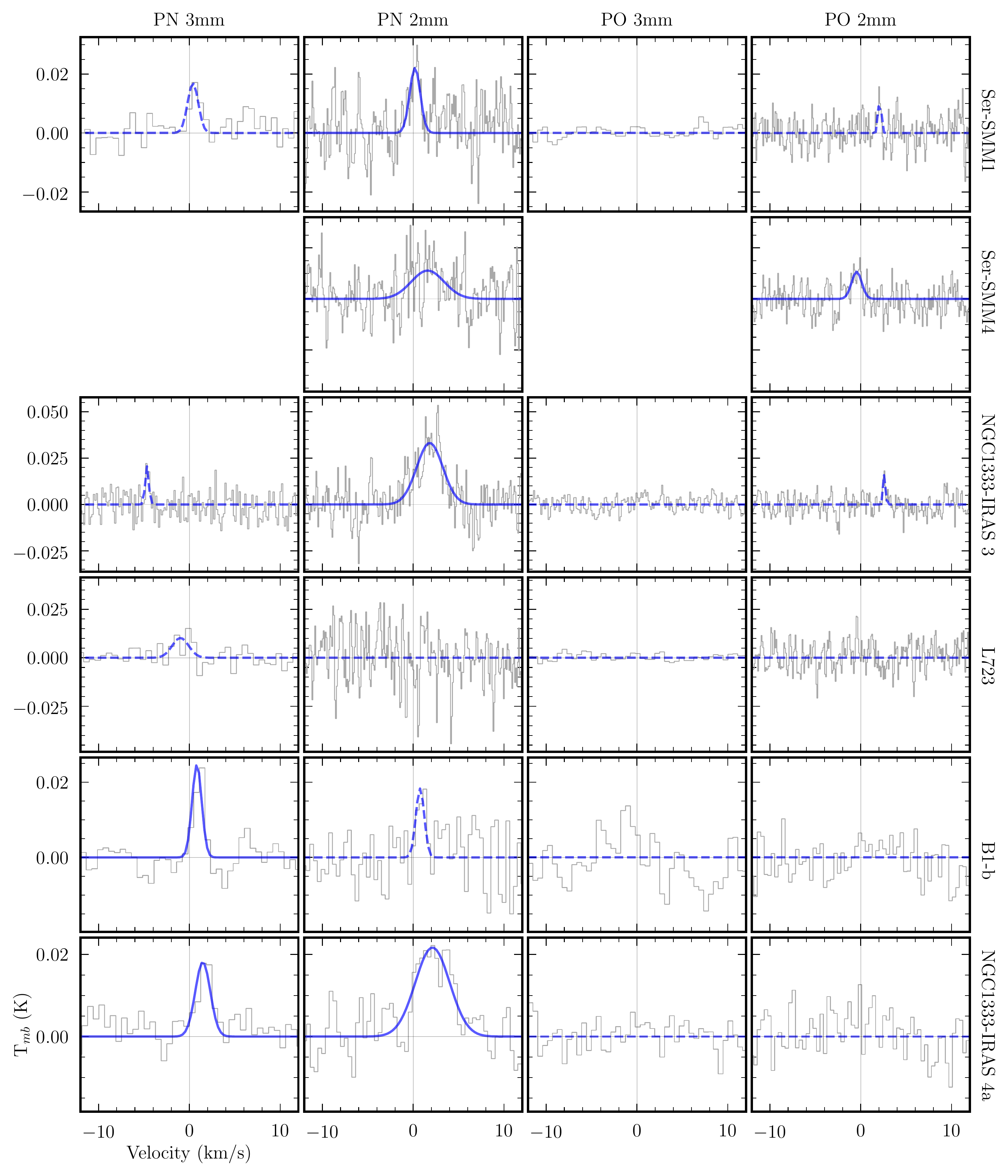}}
    \caption{PN and PO 2mm and 3mm spectra for sources with at least one detection.  The PO lines shown here are the result of stacking four hyperfine PO components. Observed spectra are shown in gray and Gaussian fits are shown in navy. For tentative detections (2--3$\sigma$), the Gaussian fits are shown with dashed lines.  Non-detected lines show a flat dashed line. Ser-SMM4 was observed only in the 2mm setup. Velocities are centered on the rest velocity of each source (Table \ref{tab:targets}).}
    \label{fig:fig1}
\end{center}
\end{figure*}

We consider a line to be firmly or tentatively detected if the velocity-integrated intensity is $>$3$\sigma$ or 2--3$\sigma$, respectively.  Figure \ref{fig:fig1} shows the observed 3mm and 2mm PO and PN spectra along with the Gaussian fits for sources with firm or tentative detections.  Note that several of the 2mm PN lines appear to be composed of multiple velocity components, however due to the low SNR we fit only a single Gaussian to each feature.  Spectra for the sources with no PO or PN detections, as well as 1.3mm spectra for the ASAI sources, are shown in Appendix \ref{app:nondets}.  

We firmly detected at least one PN line towards five sources: NGC1333-IRAS 3, Ser-SMM4, Ser-SMM1, NGC1333-IRAS 4a, and B1-b.  PN is also tentatively detected towards L723 (2.7$\sigma$), though follow-up observations are needed to confirm if this feature is real.  PO is firmly detected at 2mm in Ser-SMM4, and tentatively in Ser-SMM1 and NGC1333-IRAS 3. These tentative PO lines exhibit similar velocity offsets and line-widths as narrow features that are seen in the PN 2mm line profiles, and we therefore expect that they are not spurious.  This can be more clearly seen in Figure \ref{fig:fig3}, in which the PN and PO spectra are overlaid.  The 1.3mm PO line is also tentatively detected in B1-b (Figure \ref{fig:1mm_detections}).

Note that for NGC1333-IRAS 3 and L723, PN is only detected in one of the two observed setups. While a non-detection of the PN 2mm feature in L723 is readily explained by a low excitation temperature, it is surprising to detect the 2mm feature but not the 3mm feature in NGC1333-IRAS 3.  However, there are no other plausible line carriers nearby the 2mm PN line frequency, and we have ruled out that ghost lines could be contributing.  Moreover, the narrow spike in the PN 2mm line near 2.5 km s$^{-1}$ overlaps precisely with the PO 2mm feature, as is also seen in Ser-SMM1.  We therefore treat the 2mm PN line as a detection, and discuss the excitation conditions implied by the non-detected 3mm PN line in Section \ref{sec:nonlte}.  The PO 3mm lines in Ser-SMM1 and NGC1333-IRAS 3 are also not detected despite tentative detections in the 2mm setup.  This is likely due to the very narrow PO linewidths and low SNRs, combined with the coarse (200 kHz) spectral resolution of the 3mm setting.  Similarly, PO is tentatively detected at 1mm in B1-b, but not detected in the 2mm or 3mm settings.  Again, this could be due to a non-LTE excitation and/or high noise levels in the lower-frequency settings.

\subsection{Column Density Calculations} 
\label{subsec:columns}

To calculate the column densities of our detected molecules, we followed the population diagram method laid out by \citet{Goldsmith99}. This method assumes that the observed transitions are optically thin and well described by local thermodynamic equilibrium (LTE). We discuss the effects of non-LTE conditions in Section \ref{sec:nonlte}.

For a given transition, we find the upper-state population (N$_u$) divided by the upper-state degeneracy (g$_u$) using the velocity-integrated main-beam temperature $\int T_\mathrm{mb} dV$:
\begin{equation}
\frac{N_u}{g_u} = \frac{8 \pi k_{b}\nu^2 \int T_\mathrm{mb} dV}{h c^3 A_{ul} g_u},
\end{equation}
where $\nu$ is the line frequency and $A_{ul}$ is the transition Einstein coefficient.  To go from the upper-level populations to the total column density (N$_T$), we then use:
\begin{equation}
\frac{N_T}{Q(T_r)} = \frac{N_u}{g_u} e^{-E_u/T_r}
\label{eq:rot_diag}
\end{equation}
where E$_u$ is the upper state energy (K), $T_r$ is the rotational temperature, and $Q(T_r)$ is the partition function. 
If the emitting region does not fill the beam, beam dilution ($\eta_{bf}$) must be accounted for. We calculate the beam dilution factor by:
\begin{equation}
\eta_{bf}  = \frac{R_{src}^2}{(\frac{\theta}{2})^2 + R_{src}^2},
\end{equation}
where $\theta$ is the beam FWHM in arcsec and R$_{src}$ is the source emitting radius.  The IRAM 30m beam FWHM is 27$\arcsec$ at 90 GHz and 16$\arcsec$ at 150 GHz.  We divide each upper-level population by the appropriate beam dilution factor before calculating the column density.

When multiple transitions of a given molecule are detected, we fit Equation \ref{eq:rot_diag} to the observed upper-level populations to derive the total column density and the rotational temperature. When only a single transition is available, we assume a rotational temperature to calculate the column density.

\subsubsection{PN and PO} 
\label{subsubsec:pn_po_columns}
In the case of Ser-SMM1, B1-b, and NGC1333-IRAS 4a, we detected two PN lines and could fit for both the rotational temperature and column density.  For PN and PO in all other sources, we detected only a single line, and solved for the column density assuming a rotational temperature of 10 K based on the values found in Ser-SMM1, B1-b, and NGC1333-IRAS 4a, and previously in the low-mass sources B1-a and L1157-B1 \citep{bergner19,Lefloch2016}.  Because the source size is unknown, and cannot be determined from only one or two transitions, we adopt an emission radius corresponding to a linear scale of 1000 au for each source.  This is based on imaging of PN and PO in the low-mass protostar B1-a, which shows the phosphorus molecules emit from distinct clumps with radii of this scale \citep{bergner21}. Note that the single-dish line fluxes for B1-a were fully recovered in the interferometric observations, ruling out a more diffuse emission component.  The resulting column densities are shown in Table \ref{tab:col_densities}.

When we detected no line of PN or PO in a source, we solved for the column density upper limit using the 3$\sigma$ upper limit on the integrated intensity, where $\sigma$ = RMS $\times$ FWHM /$\sqrt{n_{ch}}$, and $n_{ch}$ is the number of channels spanned by the FWHM. We found the RMS in a 30 km s$^{-1}$ spectral window around the line position, and assumed the average FWHM of CH$_3$OH lines in the same source.

If the emitting radius is in fact larger than our assumed value, then our reported column densities are over-estimates. While this introduces a large uncertainty in our derived column densities, it is not currently possible to make a better estimate since the origin of phosphorus molecules within the protostellar environment remains poorly understood.  Additionally, the beam dilution correction (Equation 3) assumes a Gaussian emission distribution, although the distributions may be more complicated in reality.   Ultimately, spatially resolved observations of PN and PO are needed to obtain more robust column densities.

\begin{deluxetable*}{l|ccccc|ccc}
\tablenum{4}
\tabletypesize{\footnotesize}
\tablecaption{Column Densities and Excitation Temperatures \label{tab:col_densities}}
\tablehead{ \colhead{Source} & \multicolumn{5}{c}{N$_T$ (cm$^{-2}$)} & \multicolumn{3}{c}{T$_r$ (K)} \\
\colhead{} & \colhead{PN (10$^{12}$)} & \colhead{PO (10$^{12}$)} & \colhead{h-CH$_3$OH (10$^{13}$)} & \colhead{c-CH$_3$OH (10$^{13}$)} & \colhead{H$_2$ (10$^{22}$)} & \colhead{PN} & \colhead{h-CH$_3$OH} & \colhead{c-CH$_3$OH}
}
\startdata
NGC1333-IRAS 3	&	1.2 [0.2]	       & \textit{0.4 [0.2]} &	45.2 [3.4] & 6.5 [1.3]  &	2.9 [0.7] &	10	      	&	75.2 [4.9]		&	6.7 [1.4]		\\
Ser-SMM1	    &	1.7 [1.3]	       & \textit{1.0 [0.4]} &	51.4 [4.0] & 39.5 [5.9] &	9.4 [1.6] &	5.1 [1.7]   &	67.0 [8.3]		&	5.3 [0.4]		\\
L723	        &	\textit{0.8 [0.3]} & $<$1.0	           	&	7.2 [1.6]  & 8.1 [1.1]  &	1.2 [0.3] &	10		    &	62.0 [14.8]		&	6.8 [0.9]		\\
Ser-SMM4     	&	1.3 [0.4]          &  2.9 [0.9]	       	&	--	       & 34.3 [1.7] &	--	      &	10	        &	--	        	&	22.9 [1.5]	  \\
L1527	        &	$<$0.05             & $<$0.2	           	&	--	       & 3.9 [0.7]  &	2.2 [0.2] &	10	        &	--	        	&	5.8 [0.8]		\\
L1551-IRS5	    &	$<$0.1             & $<$0.4	           	&	--	       & 6.8 [0.9]  &	3.5 [0.7] &	10	        &	--	        	&	7.1 [1.2]		\\
Ser-SMM3	    &	$<$0.6             &  $<$2.0	       	&	--	       & 14.6 [0.8] &	--	      &	10	        &	--	        	&	22.5 [1.2]	  \\
NGC1333-IRAS 4a	&	0.9 [0.4]	& 	$<$1.0	          &	77.0 [7.8]  &	57.7 [9.6]	&	--	& 13.5 [7.4]	&	59.1 [14.1]	&	11.9 [1.8]	\\
B1-b	&	1.7 [1.2]	& \textit{3.8 [1.5]}&	10.0 [2.5]	&	21.0 [3.0]	&	--	&  3.4 [1.0]	&	57.7 [16.2]	&	5.7 [0.5]	\\
\enddata
\tablecomments{1$\sigma$ uncertainties are listed in brackets.  Tentative (2-3$\sigma$) detections are listed in italics, and upper limits are 3$\sigma$. When a two component fit was used to derive the CH$_3$OH column density, the hot (h) and cold (c) column densities and rotational temperatures are listed separately; single-component fits are listed as cold CH$_3$OH.  PN and PO column densities are calculated for an emitting radius of 1000 au, and with an assumed rotational        temperature of 10 K unless listed otherwise.  Note that the PN and PO column densities for NGC1333-IRAS 3 are particularly uncertain due to likely non-LTE effects (Section \ref{sec:nonlte}).}
\end{deluxetable*}

\subsubsection{CH$_3$OH}
Our observations also covered multiple CH$_3$OH lines in each of the 2mm and 3mm setups.  To derive column densities, we fitted rotational diagrams as described in Section \ref{subsec:columns}.  We used exclusively transitions of the E isomer of CH$_3$OH in calculating column densities, since our observations did not cover enough bright A-CH$_3$OH transitions to independently fit rotational diagrams for this isomer.  We then solved for the total CH$_3$OH column density assuming a CH$_3$OH E/A ratio of 0.5 based on previous surveys of CH$_3$OH towards protostellar outflow sources \citep{Holdship2019}.  

Given the larger number of CH$_3$OH transitions detected compared to the phosphorus molecules, our fitting approach differed in several ways.  First, we attempted to fit for the source size in addition to the column density and rotational temperature, but found there was no preference for a small source size.  Given this, we assumed that the CH$_3$OH emission fills the beam for subsequent fitting.  Note that JCMT observations of similar sources (including the Serpens sources in our sample) reveal large-scale CH$_3$OH emission (tens of au) surrounding each protostellar core \citep{Kristensen2010}.  Therefore, it is reasonable that CH$_3$OH emission would fill the IRAM 30m beam for our observations (16-27'').  Second, we found that some sources required both a hot and cold component to fit the observed upper level populations, as described in detail in Appendix \ref{app:ch3oh}.  Table \ref{tab:col_densities} shows the column densities and excitation temperatures of both components when relevant.  Finally, given the larger number of data points, we performed the rotational diagram fitting using the Markov Chain Monte Carlo code \texttt{emcee} \citep{Foreman2013} as this provides more robust estimates of the fit uncertainties.  CH$_3$OH rotational diagrams are shown in Appendix \ref{app:ch3oh}.

\subsubsection{C$^{17}$O and H$_2$}
For sources observed in the 3mm setup, our observations covered the C$^{17}$O 1--0 hyperfine complex at 112 GHz.  To constrain the C$^{17}$O column density in these sources, we fit the C$^{17}$O hyperfine spectrum following the procedure described in \citet{bergner19}.  The C$^{17}$O spectra are shown in Appendix \ref{app:ch3oh}.  We assume the C$^{17}$O emission fills the beam since CO should be present throughout the protostellar envelope.  The derived column densities are therefore beam-averaged.  The C$^{17}$O column density was then used to estimate the H$_2$ column density by assuming a CO/C$^{17}$O ratio of 2005 \citep{Wilson1999}, and a H$_2$/CO ratio of 10$^4$.  The resulting H$_2$ column densities are listed in Table \ref{tab:col_densities}.

\subsection{Non-LTE Effects}
\label{sec:nonlte}
As noted in Section \ref{sec:linefits}, the non-detection of the 3mm PN line towards NGC1333-IRAS 3 is not consistent with LTE emission: for any positive excitation temperature, the 3mm line should be detected above the observed noise levels.  To explore whether non-LTE PN emission could explain this, we ran a grid of \texttt{RADEX} models \citep{vanderTak2007}, using the PN collisional rates from the LAMDA database \citep{Schoier2005} with data taken from \citet{Tobola2007}.  

\begin{figure}[t]
\begin{center}
	{\includegraphics[width=1\linewidth]{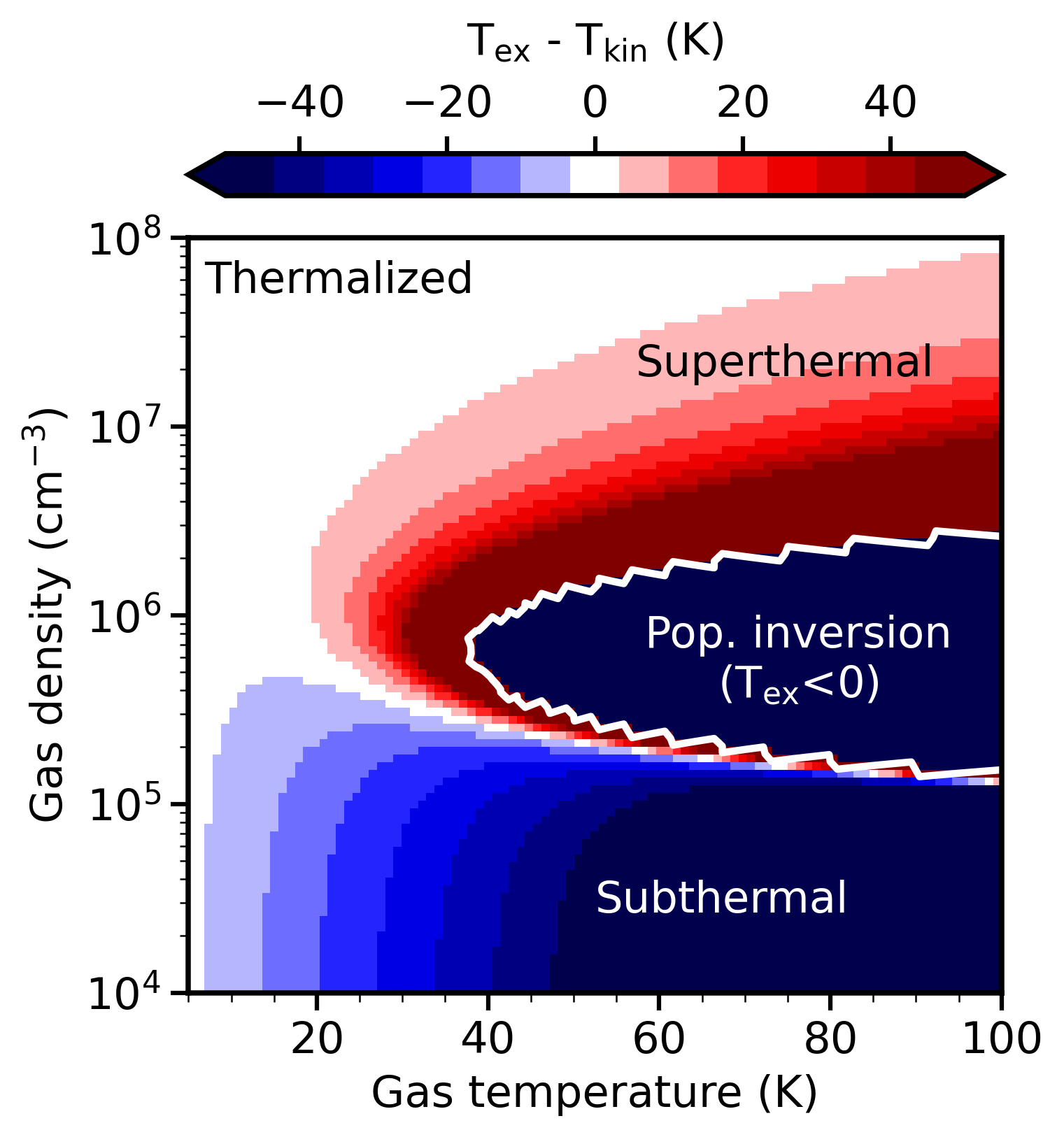}}
    \caption{Difference between the \texttt{RADEX}-calculated excitation temperature of the PN 2--1 line and the gas kinetic temperature, shown for a range of gas densities and temperatures.  Different excitation regimes are indicated: the line is thermalized in the white region, superthermally excited in the red region, and subthermally excited in the blue region.  Interior to the white contour, T$_\mathrm{ex}$ is $<$0, reflecting a population inversion ($n_u/g_u > n_l/g_l$).}
    \label{fig:pn_excitation}
\end{center}
\end{figure}

Figure \ref{fig:pn_excitation} shows the excitation landscape for the PN 2--1 (3mm) line, assuming a line width of 1 km s$^{-1}$ and a column density of 10$^{12}$ cm$^{-2}$.  For low gas temperatures and high gas densities, the line is thermalized.  At high temperatures and high densities, the line emits superthermally, while at low densities and all temperatures it emits subthermally.  Importantly, for gas densities around 10$^6$ cm$^{-3}$ and temperatures $>$40 K, the 2--1 line is characterized by a negative excitation temperature, i.e.~a population inversion.  The PN 3--2 line shows a similar overall excitation map as the 2--1 line, but the population inversion region is shifted to higher temperatures ($>$70 K).  

Given that phosphorus molecule emission is associated with outflow shocks, it is certainly plausible that the emitting region of NGC1333-IRAS 3 would be warm and dense.  Thus, we expect that the non-detection of the PN 3mm line can be explained by a level population inversion.  That this behavior is not seen for other sources in our survey, or for previously observed low-mass protostars \citep{Lefloch2016, bergner19} may reflect that the emitting gas is warmer and/or denser in NGC1333-IRAS 3.  Indeed, with a non-LTE analysis of four PN lines, \citet{Lefloch2016} infer a gas density of $\sim$10$^5$ cm$^{-2}$ and a kinetic temperature around 40--80 K, i.e.~the `subthermal' regime of Figure \ref{fig:pn_excitation}.  The low (but positive) PN excitation temperatures found in Ser-SMM1, B1-b, and NGC1333-IRAS 4a also point to emission either within the subthermal or thermalized regimes, corresponding to lower densities or lower temperatures compared to the population inversion regime.

The complex excitation behavior of PN illustrated in Figure \ref{fig:pn_excitation} underscores the need for multi-line coverage to obtain good constraints on the column densities and emitting conditions.  Indeed, with only two lines it is not possible to perform a full non-LTE fit with our current observations, and coverage of higher-J PN lines is needed to fully understand the excitation of the emitting molecules.  Still, we note that for the cool gas temperatures ($<$40 K) tested in \citet{bergner19}, the derived PN column densities are not significantly affected by non-LTE effects.

\section{Results} \label{sec:results}
\subsection{Column Densities} \label{subsec:res_columns}
The column density findings for each source are presented in Table \ref{tab:col_densities}.  For sources with detections, the column densities of both PN and PO range are around 10$^{12}$ cm$^{-2}$.  Non-detection upper limits are on the order $<$10$^{11}$--10$^{12}$ cm$^{-2}$. The phosphorus molecule column densities in our sample are about one order of magnitude lower than was reported in Ori (KL), W51M, and Sgr B2 \citep{Turner1987}, several high mass protostars.  The PO and PN column densities in our sample are in good agreement with the low-mass protostars L1157 B1 \citep{Yamaguchi2011} and B1-a \citep{bergner19}. 

Because the degree of beam dilution remains uncertain, it is important to note that our PN and PO column densities depend on the actual source size (R$_{src}$) of the emitting region. If R$_{src}$ is smaller than estimated (1000 au), the true column densities will be higher than listed, and the opposite for a larger R$_{src}$. If T$_{rot}$ is lower (e.g. 5 K rather than 10 K), the PN and PO column densities could be underestimated by a factor of 2--3.  A higher T$_{rot}$ (20 K) does not strongly affect the PN or PO column density.

CH$_3$OH column densities span $\sim$10$^{13}$--10$^{14}$ cm$^{-2}$ in our sample, and fall within the range found in a survey of embedded low-mass protostars by \citet{Graninger2016}.  H$_2$ column densities in our sample are $\sim$10$^{22}$ cm$^{-2}$, which is an order of magnitude higher than the H$_2$ column density found in B1-a \citep{bergner19}.

\subsection{Column density ratios} \label{subsec:ratios}
In Table \ref{tab:ratios}, we present the column density ratios of PO and PN relative to other molecules. The PO/PN ratio is of particular interest for comparing the phosphorus chemistry in our source sample with other star-forming regions. Intriguingly, to date all source-averaged measurements of PO/PN in dense interstellar regions are between 1 and 3, despite probing a wide range of physical environments \citep{Lefloch2016, Rivilla16, Rivilla2018, bergner19, rivilla20, Bernal21}. Across our sample, we find PO/PN ratios within this range or somewhat lower (0.6--2.2).  Note that we exclude NGC1333-IRAS 3 from this comparison, since it likely suffers from non-LTE effects for at least the PN emission (Section \ref{sec:nonlte}).  Coverage of additional PN and PO lines is needed to perform a full non-LTE fit and confirm the PO/PN ratio in NGC1333-IRAS 3.

In some sources (e.g.~Ser-SMM1, NGC1333-IRAS 3), the kinematics of the PN and PO lines are different, suggesting that they may emit from different gas.  This is consistent with previous spatially resolved observations of phosphorus molecules: in both AFGL 5142 and B1-a \citep{rivilla20, bergner21}, PN and PO emit from several distinct emission clumps which have different velocity offsets and also different PO/PN ratios.  This is likely true for these sources as well.  We emphasize that the derived PO/PN ratios are averages across the source, and that there is likely small-scale variation in the PO/PN ratio across these sources.  

We also present $\mathrm{\frac{PN+PO}{H}}$ abundances as an indicator of the total phosphorus abundance in the gas. For sources with H$_2$ column density estimates, we find $\mathrm{\frac{PN+PO}{H}}$ abundances of the order 10$^{-11}$. This is several orders of magnitude below the Solar P abundance of 3$\times$10$^{-7}$ \citep{asplund2009}, and on the low end but consistent with the range of P/H ratios previously inferred in the dense ISM \citep{Lefloch2016, Rivilla16, bergner19, Bernal21}.  Note that we use beam-averaged H$_2$ column densities to estimate this ratio.  Spatially resolved observations are needed to constrain the H$_2$ column density co-spatial with the phosphorus molecules.  

Lastly, we list the $\mathrm{\frac{PN+PO}{CH_{3}OH}}$ ratio to enable a comparison with cometary ice compositions (Section \ref{subsec:comets}). Because CH$_3$OH does not form efficiently in the gas, the gas-phase CH$_3$OH content in these sources should originate from ice desorption. Assuming that phosphorus is released into the gas from the same sputtering process, the observed gas-phase $\mathrm{\frac{P}{CH_{3}OH}}$ ratio should reflect the underlying ice-phase $\mathrm{\frac{P}{CH_{3}OH}}$ ratio.  For sources with both a hot and cold CH$_3$OH component (Table \ref{tab:col_densities}), we use only the cold component to find this ratio given that the excitation conditions are closer to the phosphorus molecules.  We find $\mathrm{\frac{PN+PO}{CH_{3}OH}}$ ratios of $\sim$0.7--2.7\% in our sample, with a low upper limit ($<$0.3\%) found for NGC1333-IRAS 4a.

\begin{deluxetable}{l|ccc}
\tablenum{5}
\tablecaption{Column Density Ratios\label{tab:ratios}}
\tablehead{\colhead{Source} & \colhead{$\mathrm{\frac{PO}{PN}}$} & \colhead{$\mathrm{\frac{PN+PO}{H}}$} &  \colhead{$\mathrm{\frac{PN+PO}{CH_{3}OH}}$} \\
& & \colhead{(10$^{-11}$)} &  \colhead{(\%)}
}
\startdata
NGC1333-IRAS 3$^*$	&	0.3 [0.1] 	&	2.7 [0.7] &  2.4 [0.6]\\
Ser-SMM1	&	 0.6 [0.5]	&	1.5 [0.8]	 &	0.7 [0.4]	\\
Ser-SMM4	&  2.2 [1.0] &	-	& 1.2 [0.3]\\
B1-b        &  2.2 [1.7] & -   &  2.7 [1.0]\\
NGC1333-IRAS 4a    & $<$1.1   & - &$<$0.3\\
\enddata
\tablenotetext{}{$^*$Note that the values for NGC1333-IRAS 3 are particularly uncertain due to likely non-LTE effects (Section \ref{sec:nonlte}).}
\end{deluxetable}

\begin{figure*}[t]
\begin{center}
	{\includegraphics[width=\textwidth]{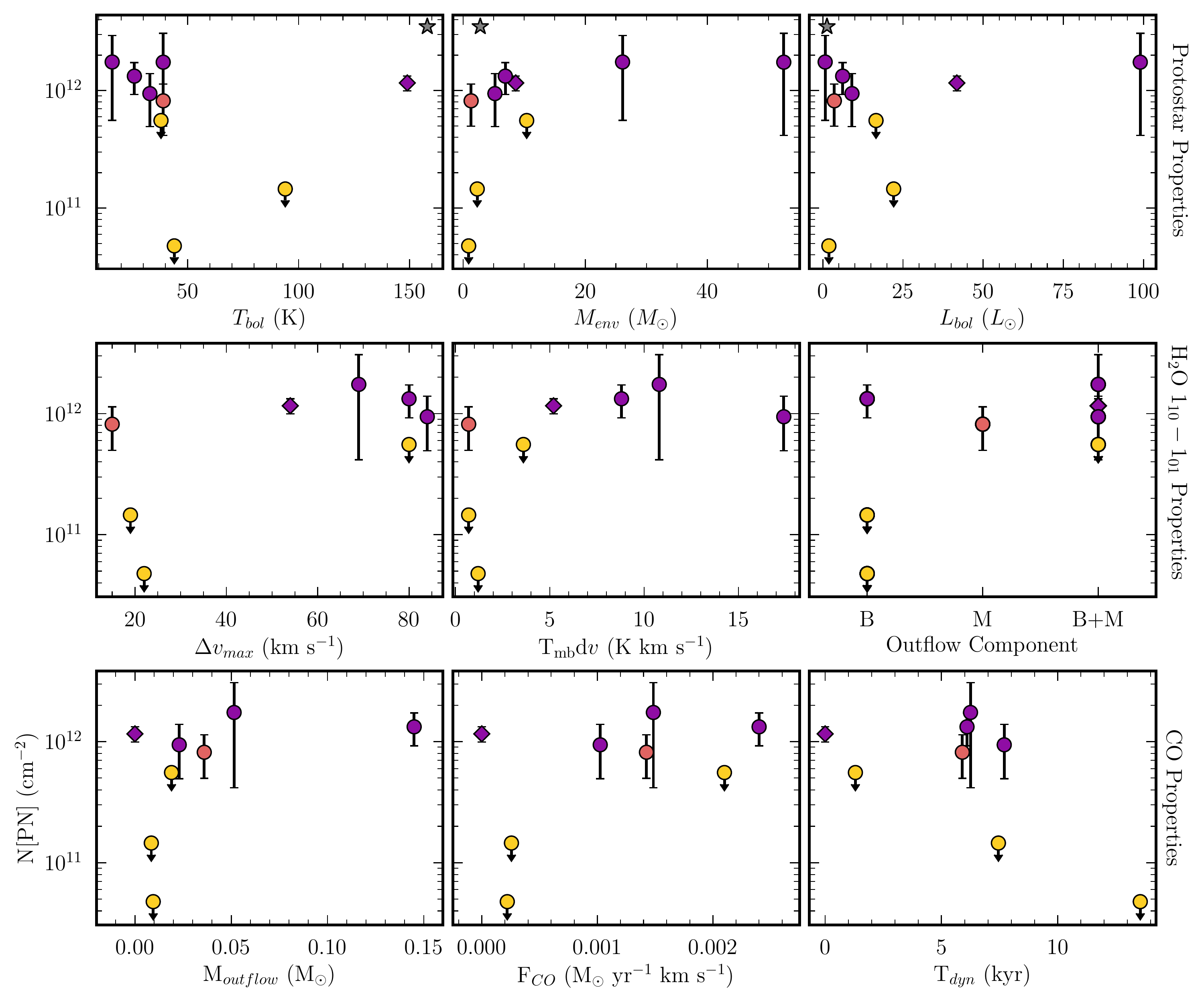}}
    \caption{PN column densities plotted against the protostar physical properties (top row; Table \ref{tab:targets}), H$_2$O 1$_{10}$--1$_{01}$ emission properties \citep[middle row;][]{kristensen12}, and CO-derived outflow properties \citep[bottom row;][]{Yildiz2015}, described in detail in the text. Purple represents a firm detection, pink represents a tentative (2--3$\sigma$) detection, and yellow represents an upper limit non-detection. B1-a is shown as a grey star.  Note that the NGC1333-IRAS 3 column density,  plotted as a diamond, is particularly uncertain due to non-LTE effects (Section \ref{sec:nonlte}).}
    \label{fig:fig2}
\end{center}
\end{figure*}

\subsection{Trends across the source sample}
\label{subsec:sample_trends}
With five firm detections of phosphorus molecules towards low-mass star forming regions, we can now search for relationships between the source properties and the gas-phase phosphorus inventory. Our source sample was selected to span a range of physical properties (e.g.~envelope mass) and evolutionary stages (traced by bolometric temperature).  Additionally, most sources in our sample were part of the Herschel WISH program and have auxiliary constraints on the outflow characteristics, traced by H$_2$O and CO emission \citep{kristensen12, Yildiz2015}.  In general, we expect the H$_2$O emission to correlate with shocked gas, whereas CO emission traces the bulk mechanical properties of the outflows.

Figure \ref{fig:fig2} shows the the PN column density as a function of the protostar and outflow properties.  We focus on PN since it is firmly detected towards more sources than PO.  Comparisons between the PN column density and the protostar properties (top row) reveal no trends.  PN is detected towards both Class 0 (T$_{bol}<70$K) and Class I (T$_{bol}>$70K) sources, and towards sources with envelope masses ranging from 3--52 M$_\odot$ and bolometric luminosities from 1--100 L$_\odot$. Thus, the release of phosphorus into the gas does not appear correlated with any intrinsic physical property of the protostar.

The middle row of Figure \ref{fig:fig2} shows comparisons between the PN column density and various features of the H$_2$O 1$_{10}$--1$_{01}$ emission: the line full-width at zero-intensity ($\Delta v_{max}$), velocity-integrated line intensity (T$_\mathrm{mb}$d$v$), and kinematic components (broad, medium, or both, as defined in Section \ref{subsec:sources}), all taken from \citet{kristensen12}. While we do not see any monotonic relationships, we do find similarities among the sources where PN is detected.  The sources where PN is firmly detected have the highest velocity-integrated H$_2$O intensities of the sample ($>$5 K km s$^{-1}$).  These same sources also have broad H$_2$O linewidths ($\Delta v_{max}>$50 km s$^{-1}$). Together, this indicates that phosphorus is released into the gas when outflow shocking is relatively strong, producing broad H$_2$O lines and a high H$_2$O content in the gas. 
 
Figure \ref{fig:fig2} (bottom) shows comparisons with the outflow mass (M$_{outflow}$), outflow force (F$_{CO}$), and dynamical age (T$_{dyn}$) measured from CO 3--2 emission in \citet{Yildiz2015}.  Note that we adopt the average values between the red and blue outflow lobes.  In all cases, we do not see any correlations: PN is both detected and not detected over a range of outflow masses, forces, and ages.  This lack of correlation with CO-derived mechanical properties, taken together with the association between PN and H$_2$O emission, implies that gas-phase phosphorus is specifically related to shocked gas, rather than the bulk outflowing material.

We also attempted to look for trends between the source or outflow properties and the PO/PN ratio.  However, given the narrow range in PO/PN ratios across the sample (Table \ref{tab:ratios}) it is not possible to identify any clear relationships with this sample.

\subsection{Line profile comparisons}
\label{subsec:lineprofs}
\begin{figure*}[t]
\begin{center}
	{\includegraphics[width=\textwidth]{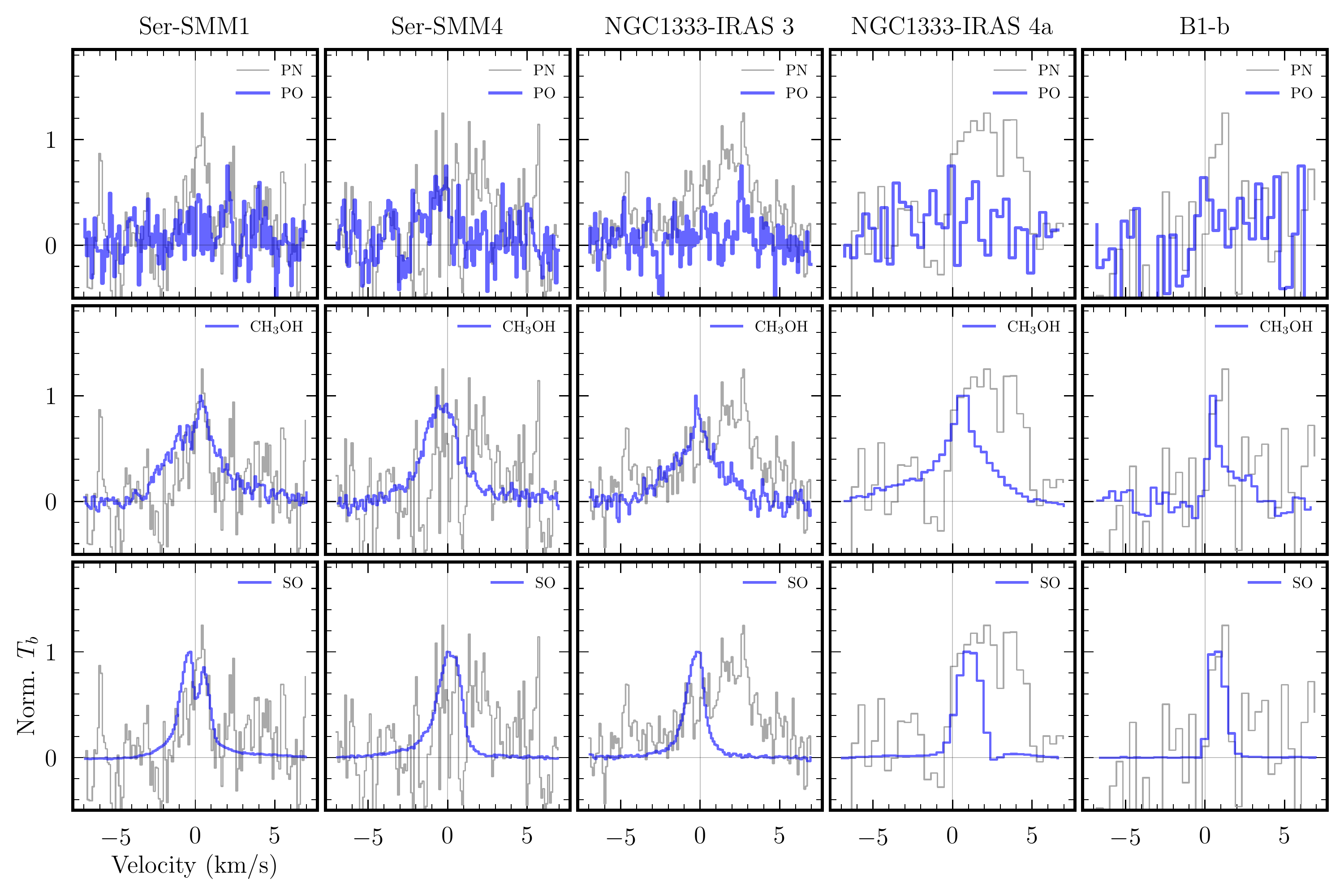}}
    \caption{Top: normalized line profiles of the 2mm PO line (E$_u$=15.7 K) and PN line (E$_u$=13.5 K) for each source. Middle and bottom: normalized line profiles of the 2mm PN lines compared to CH$_3$OH (157.246 GHz, E$_u$ = 36K) and SO  (99.30 GHz, E$_u$= 16K). Velocities are centered on the rest velocity of each source (Table \ref{tab:targets}).}
    \label{fig:fig3}
\end{center}
\end{figure*}

In Section \ref{subsec:sample_trends}, we showed that the detection of phosphorus molecules appears to trend with the presence of shocked gas as traced by the H$_2$O 1$_{10}$--1$_{01}$ line.  However, the PO and PN lines in our sample are significantly narrower than the H$_2$O lines presented in \citet{kristensen12}, a few km s$^{-1}$ compared to tens of km s$^{-1}$, indicating that P molecules emit from a much narrower range of physical environments within the outflow compared to H$_2$O.   Moreover, the PO and PN lines are offset by just a few km s$^{-1}$ from the source rest velocities, reflecting an origin in either weakly shocked or post-shocked gas.  To further explore the origin of phosphorus molecule emission within the protostellar environment, Figure \ref{fig:fig3} shows the line profiles of PN and PO with one another and with the weak-shock tracers CH$_3$OH and SO.

In both Ser-SMM1 and NGC1333-IRAS 3, the PO line is a narrow feature red-shifted by $\sim$3 km s$^{-1}$ relative to the source rest velocity.  The PN lines in both sources exhibit a broader component red-shifted by $\sim$1--2 km s$^{-1}$, along with a narrow component with a similar width and offset to the PO feature.  This may reflect that PO preferentially emits from higher-velocity material within the shock, while PN emits from multiple environments.  In Ser-SMM4, a similar pattern is seen, in which the PO line is narrower but overlaps with the PN feature.  In this case, the PO line is much broader than those seen in Ser-SMM1 and NGC1333-IRAS 3, and the PO line is blue-shifted rather than red-shifted relative to the source velocity.

We also compare the PN line profiles to the weak-shock tracers CH$_3$OH and SO.  In the low-mass protostar B1-a, the kinematics of CH$_3$OH and the sulfur carrier SO$_2$ were found to closely resemble the phosphorus molecules \citep{bergner19}, and similarities between PN and SO abundances have been previously identified in numerous massive star-forming regions \citep{Mininni2018, rivilla20}.  Among our sources, the CH$_3$OH and SO lines generally peak close to the source rest velocity, with SO characterized by a narrower line width compared to CH$_3$OH.  Ser-SMM1, Ser-SMM4, and B1-b exhibit similarities in the line shapes and velocity offsets of PN compared to CH$_3$OH and SO.  This is particularly true when comparing SO and PN in Ser-SMM1 and B1-b.  The PN line profiles do not agree well with CH$_3$OH or SO in NGC1333-IRAS 3 and NGC1333-IRAS 4a.  In both cases, there is a strong PN emission component red-shifted with respect to the CH$_3$OH and SO lines.  Thus, while there is some overlap, neither CH$_3$OH nor SO appear to uniquely trace the same material as the phosphorus molecules in these sources.

\section{Discussion} \label{sec:discussion}
Our survey, combined with archival observations \citep{Lefloch2018}, has enabled us to characterize the emission of phosphorus molecules towards five new Solar-type star forming regions. Here, we discuss the implications of our findings with respect to the interstellar phosphorus chemistry and the volatile phosphorus reservoir in the young Solar System.

\subsection{Phosphorus molecules and outflow properties}
\label{subsec:disc_origins}
Prior detections of PO and PN in the low-mass star forming regions B1-a \citep{bergner19} and L1157-B1 \citep{Yamaguchi2011, Lefloch2016} suggest that the phosphorus emission in those sources originates from outflow shocks.  However, the exact relationship between the phosphorus chemistry and outflow physics currently remains ambiguous.

Of the sources targeted in this work, we firmly detect phosphorus molecule emission only towards sources with the highest H$_2$O 1$_{10}$-1$_{01}$ fluxes (Figure \ref{fig:fig2}), reflecting robust shock-induced sputtering of icy material from the grains. We do not identify any correlations between phosphorus molecules and either the protostellar properties or the bulk mechanical properties of the outflowing gas.  This implies a particular association between gas-phase phosphorus and shocked gas.  This is consistent with the emission patterns seen in the low-mass protostar B1-a, in which PN and PO emit not from the SiO outflow, but from distinct shock positions associated with ice sublimation \citep{bergner21}.  

The PO and PN lines are characterized by narrow linewidths ($<$2 km s$^{-1}$) and small offsets from the source systemic velocities (a few km s$^{-1}$).  This indicates that the phosphorus molecules do not originate in strongly shocked material itself, but rather in a post-shocked stage. In this scenario, shocking is required to release the parent phosphorus carriers from the surface of grains, followed by gas-phase chemistry in the post-shocked gas to form PO and PN.  Indeed, \cite{rivilla20} and \citet{bergner21} infer a similar origin in post-shocked gas based on spatially resolved observations of PO and PN towards a high-mass and low-mass protostar, respectively.  Notably, however, the PN line in L1157-B1 extends to velocities as high as 20 km s$^{-1}$ \citep{Lefloch2016}, suggestive that the emission in this source may be more closely connected to the shocked material.

In our sources, we see interesting velocity structure in the phosphorus molecule line profiles (Figure \ref{fig:fig3}).  In Ser-SMM 4, NGC1333-IRAS 3, and NGC1333-IRAS 4a, the PN lines are fairly broad and appear to be composed of several velocity components.  In Ser-SMM1 and NGC1333-IRAS 3, PO emits as a very narrow feature a few km s$^{-1}$ offset from the main PN line, coincident with a very narrow PN feature.  This velocity structure likely corresponds to emission from different shock positions along the outflow.  Indeed, spatially resolved observations towards both AFGL 5142 and B1-a show that PO and PN emit from several distinct clumps adjacent to the outflow, which are characterized by different velocity offsets \citep{rivilla20, bergner21}.  That PO generally exhibits fewer velocity components compared to PN in our sources suggests that it emits from a smaller number of clumps along the outflow.  This is analogous to the emission pattern seen in B1-a, where PN emits brightly from two emission clumps, while PO emits brightly from only one clump \citep{bergner21}.  
Resolved observations are needed to further explore the the progression of post-shock chemistry in these sources, and the origins of chemical differentiation at different shock positions.

\subsection{Volatile P in low-mass protostars} \label{subsec:comparrison}

In the two sources for which we could measure the H$_2$ column density, we estimate source-averaged $\mathrm{\frac{PN+PO}{H}}$ ratios of 1.5--2.7 $\times$10$^{-11}$ (Table \ref{tab:ratios}).  This corresponds to $<$0.01\% of the Solar P abundance \citep{asplund2009}. For reference, the combined PO and PN abundances in L1157-B1 and B1-a are 3.6$\times$10$^{-9}$ and 10$^{-10}$--10$^{-9}$, respectively.  Thus, our measured abundances support previous findings that volatile P is only a trace component of the total P inventory in planetary system progenitors.  It is important to note that our PO and PN column densities are fairly uncertain due to unknown beam dilution factors.  Still, even adopting a very small emitting area of 1$\arcsec$ in radius would only increase the derived abundances by an order of magnitude, leading to an abundance that is still significantly lower than the Solar P/H ratio.

The column density ratios of PO/PN in our sample range from 0.6 to 2.2 (Section \ref{subsec:ratios}). These values are generally consistent with the source-averaged PO/PN ratios reported in the other low-mass protostars B1-a and L1157-B1, which range from 1 to 3 \citep{Lefloch2016, bergner19}.  Shock chemistry models predict that PO/PN should be sensitive to factors like the pre-shock duration, shock density, shock velocity, time post-shock, and gas-phase O/N ratio \citep{Aota2012, Lefloch2016, Jimenez-Serra2018}.  However, we find a small range in source-averaged PO/PN ratios across our sample: a factor of 3.5, as compared to 1-2 orders of magnitude of variations in the source and outflow properties (Section \ref{subsec:sources}).  Thus, the source-averaged PO/PN ratios appear only weakly sensitive to environmental factors.

Spatially resolved observations will likely provide a clearer path to determining environmental drivers of the phosphorus chemistry: existing studies have revealed a wider range of local PO/PN ratios from $\sim$1--8 \citep{rivilla20,bergner21}.  Follow-up mapping towards the sources with PN and PO detections would provide valuable new constraints on how and why the phosphorus chemistry varies within a protostellar environment.

\subsection{Comparison to Comet 67P}
\label{subsec:comets}
Comets represent the most minimally processed relics from the planet formation epoch in our Solar system.  Volatile P was recently detected in the Solar System comet 67P/Churyumov–Gerasimenko \citep{Altwegg16}, with a reported PO/PN ratio of at least 10 \citep{rivilla20}.  The source-averaged PO/PN ratios previously measured towards the proto-Solar analogs L1157 and B1-a are low ($<$3) compared to comet 67P \citep{Lefloch2016, bergner19}.  With our expanded sample, we have shown that B1-a and L1157 are not unusual in their low PO/PN ratios, and that source-averaged PO/PN ratios in low-mass star forming regions appear in general to be lower than the cometary measurement.  This difference between the protostellar and cometary PO/PN ratios may be explained if the solar system exhibits an unusual volatile P chemistry compared to typical low-mass protostars, or if processing between the protostellar and disk stages alters the volatile P composition in the ice.  

The (PO+PN)/CH$_3$OH ratio can be used to compare the volatile P content between low-mass protostars and comet 67P, assuming that the observed CH$_3$OH, PO, and PN emission all trace material sputtered from the grains by shocks.  In our sources, we find source-averaged (PO+PN)/CH$_3$OH ratios of 0.7--2.7\%, as well as a low upper limit ($<$0.3\%) for NGC1333-IRAS 4a (Table \ref{tab:ratios}).  The ratio in B1-a is typically 1--3\% \citep{bergner21}, and comet 67P was measured to have a (PO+PN)/CH$_3$OH ratio of $\sim$5\% \citep{Rubin2019}.  Thus, most measurements of (PO+PN)/CH$_3$OH along the low-mass star formation sequence are on the order of a few percent. This suggests that the volatile P content of the young Solar system is fairly typical.  Still, there is currently only one comet for which P/CH$_3$OH ratios have been measured, and better constraints on the volatile P inventory of primitive icy material in the Solar System would improve this comparison. 

To date, it is unknown whether prebiotically accessible P on the early Earth was sourced from a soluble mineral like schreibersite \citep[e.g.][]{Pasek2005}, or from the delivery of volatile icy material through impact \citep[e.g.][]{Rubin2019a, rivilla20}.  The similarity in volatile P content between a Solar System comet and protosolar analogs implies that, if impact delivery was indeed an important source of P for origins of life chemistry on Earth, then planets in other systems could also be seeded with prebiotic P in the same way.

\section{Conclusions}
We have targeted PN and PO lines towards seven low-mass protostars with the IRAM 30m telescope in an effort to expand our understanding of P astrochemistry in proto-Solar analogs.  We report new detections of phosphorus molecules towards NGC1333-IRAS 3, Ser-SMM4, Ser-SMM1.  Combined with archival observations of NGC1333-IRAS 4a and B1-b \citep{Lefloch2018}, we have characterized the PN and PO column densities and line kinematics towards five low-mass protostars, enabling explorations of the demographics of phosphorus chemistry in proto-Solar analogs. 

The sources with PN detections show evidence for powerful outflow shocks based on their H$_2$O 1$_{10}$-1$_{01}$ emission, supporting the central role of shocks in releasing P into the gas in dense SFRs.  Correlations are not found with the protostar properties or the bulk mechanical properties of the outflow, implying a specific association between gas-phase phosphorus and shocked gas.  Based on line kinematics, the PO and PN emission seems to arise from quiescent post-shocked gas rather than directly from the strongly shocked gas.  Ultimately, spatially resolved observations are required in order to unambiguously determine the origin of the P-bearing species in these protostellar regions.

While our source sample spans a wide range of protostellar and outflow characteristics, we find a fairly narrow range of source-averaged PO/PN ratios (0.6-2.2).  This suggests that the phosphorus chemistry is not particularly sensitive to environmental factors.  However, we expect the PO/PN ratios to vary locally across these sources, and spatially resolved observations are needed to further explore the drivers of this phosphorus chemistry.

With this analysis, we can begin to compare how the `typical' protostellar phosphorus chemistry compares to the young Solar System. We find that the volatile P content in our sources is consistent with that inferred for comet 67P, with (PN+PO)/CH$_3$OH ratios around 1--3\%.  However, the source-averaged PO/PN ratios in low-mass protostars are universally lower than the ratio $>$10 found in the comet.  Spatially resolved observations of additional Solar-type star forming regions, and further constraints on volatile P chemistry within the Solar System, are needed to further constrain how this biocritical element is inherited to the planet-formation epoch.

\acknowledgments
The authors are grateful to the anonymous referee for helpful feedback that improved the quality of this manuscript.  This work is based on observations carried out under project number 131-19 with the IRAM 30m telescope. IRAM is supported by INSU/CNRS (France), MPG (Germany) and IGN (Spain).  J.B.B. acknowledges support from NASA through the NASA Hubble Fellowship grant \#HST-HF2-51429.001-A awarded by the Space Telescope Science Institute, which is operated by the Association of Universities for Research in Astronomy, Incorporated, under NASA contract NAS5-26555. 

\appendix
\label{sec:app}

\section{Observational details}
\label{app:obsdeets}
Table \ref{tab:observations} provides an overview of the observations taken at the IRAM 30m telescope, including for each source the spectral setups, time on-source, system temperature, and resulting spectral RMS.  The listed image frequencies correspond the following spectral coverages: 110 GHz: 92.02--99.80 and 107.70--115.48 GHz; 109.5 GHz: 92.86--94.68, 96.14--97.96, 108.54--110.36, and 111.82--113.64 GHz; 141 GHz: 136.86--138.68, 140.14--141.96, 152.54--154.36, and 155.82--157.64 GHz.  All observations were performed in position switching mode, with the `off' positions listed in Table \ref{tab:observations}.  Pointing corrections were performed every 2h, and focus corrections every 4h.  The primary line targets are listed in Table \ref{tab:line_targets}.

\begin{deluxetable}{l|ccccccc}
\tablenum{6}
\tablecaption{Observational details\label{tab:observations}}
\tablewidth{0pt}
\tablehead{ \colhead{Source} & \colhead{Off pos.} & \colhead{Setup} & \colhead{Image freq.} & \colhead{Time on-source} & \colhead{T$_{sys}$} &  \colhead{$\tau$} & \colhead{RMS}
\\
\nocolhead{} & \colhead{($\arcsec$)} & \colhead{} & \colhead{(GHz)}& \colhead{(min.)}& \colhead{(K)} & \colhead{} & \colhead{(mK)}}

\startdata
Ser-SMM4	&	-660 500	&	2mm	&	141	&	50	&	119	&	0.029	&	9.2	\\
Ser-SMM3	&	-700 430	&	2mm	&	141	&	30	&	127	&	0.036	&	15.3	\\
Ser-SMM1	&	-550 380	&	3mm	&	110	&	82	&	107	&	0.024	&	4.1	\\
	        &	        	&	2mm	&	141	&	100	&	153	&	0.063	&	7.6	\\
L723	    &	-290 110	&	3mm	&	110	&	100	&	110	&	0.023	&	3.3	\\
        	&		        &	2mm	&	141	&	60	&	179	&	0.13	&	12.1	\\
L1527	    &	1440 580	&	3mm	&	110	&	80	&	100	&	0.022	&	4.4	\\
        	&	        	&	2mm	&	141	&	32	&	124	&	0.035	&	12.7	\\
L1551-IRS5	&	580 -450	&	3mm	&	110	&	50	&	97	&	0.018	&	4.5	\\
NGC1333-IRAS 3	&	1100 0 	&	3mm	&	109.5	&	100	&	89	&	0.035	&	7.9	\\
            &               &   2mm & 141 & 100 & 125--227 & 0.033--0.134 & 8.9 \\
\enddata
\tablecomments{Overview of new IRAM 30m observations. RMS values are centered around the PN lines at 93.98 GHz (3mm) and 140.97 GHz (2mm). NGC1333-IRAS 3 was observed over two days in the 2mm setup, leading to a range of observing conditions.}
\end{deluxetable}

\FloatBarrier
\section{PN and PO line fits and additional spectra}
\label{app:nondets}

\begin{figure*}[h]
\begin{center}
	{\includegraphics[width=\textwidth]{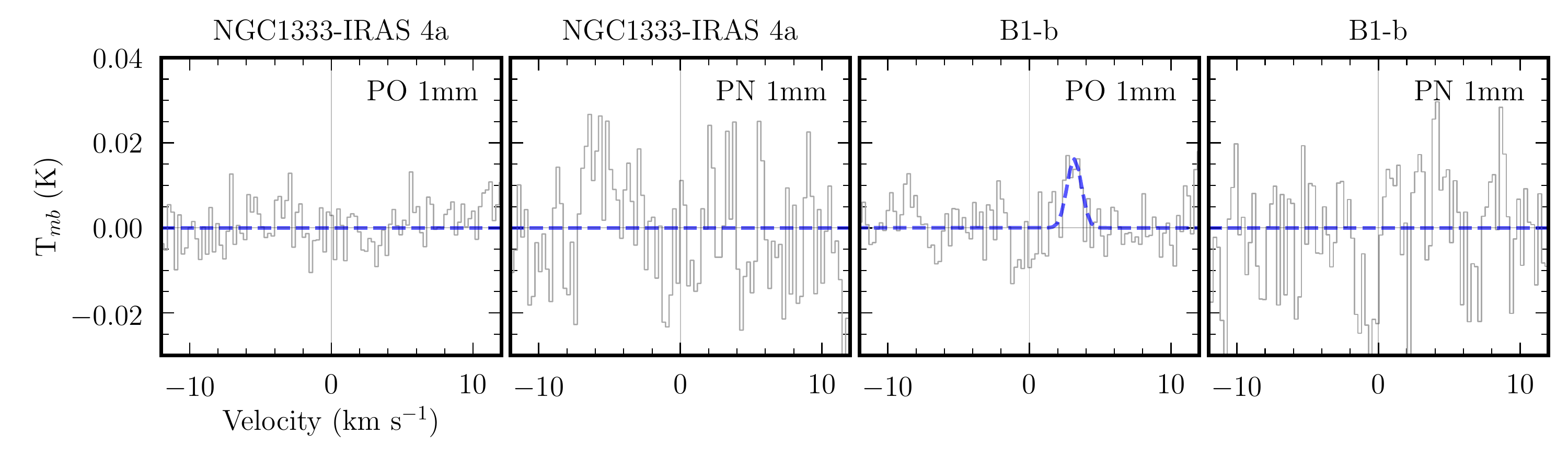}}
    \caption{As for Figure \ref{fig:fig1} but showing the 1.3mm PO and PN spectra observed towards the ASAI sources.}
    \label{fig:1mm_detections}
\end{center}
\end{figure*}

Figure \ref{fig:1mm_detections} shows the ASAI 1.3mm spectra of PN and PO, analogous to the 2mm and 3mm spectra shown in Figure \ref{fig:fig1}.  The PO line is tentatively detected towards B1-b, and the other lines are not detected.

\begin{deluxetable}{l|ccccc}
\tablenum{7}
\tablecaption{PN and PO Gaussian line fits \label{tab:p_gaussian_fitting}}
\tablehead{ \colhead{Source} & \colhead{Setup} & \colhead{Molecule} & \colhead{Integrated intensity} & \colhead{Line Width} & \colhead{Offset} \\ 
 & \colhead{} & \colhead{}  & \colhead{(mK km s$^{-1}$)} & \colhead{(km s$^{-1}$)} & \colhead{(km s$^{-1}$)}
}
\startdata
L723 	 & PN 	 & 2mm 	 & $<$17.7 	 & \textit{1.0} 	 & -- 	 \\
 	 &  	 & 3mm* 	 & 23.4 [9.1] 	 & 0.9 	 & 10.0 	 \\
 	 & PO 	 & 2mm 	 & $<$10.6 	 & \textit{1.0} 	 & -- 	 \\
 	 &  	 & 3mm 	 & $<$5.4 	 & \textit{1.0} 	 & -- 	 \\
\hline
Ser-SMM1 	 & PN 	 & 3mm* 	 & 25.0 [9.1] 	 & 0.6 	 & 8.9 	 \\
 	 &  	 & 2mm 	 & 32.7 [7.5] 	 & 0.6 	 & 8.7 	 \\
 	 & PO 	 & 3mm 	 & $<$8.4 	 & \textit{1.3} 	 & -- 	 \\
 	 &  	 & 2mm* 	 & 5.1 [2.2] 	 & 0.2 	 & 10.6 	 \\
\hline
NGC1333-IRAS 3 	 & PN 	 & 3mm* 	 & 10.2 [5.0] 	 & 0.2 	 & 3.6 	 \\
 	 &  	 & 2mm 	 & 119.5 [17.1] 	 & 1.4 	 & 10.1 	 \\
 	 & PO 	 & 3mm 	 & $<$8.1 	 & \textit{1.6} 	 & -- 	 \\
 	 &  	 & 2mm* 	 & 5.4 [2.1] 	 & 0.1 	 & 10.9 	 \\
\hline
L1527 	 & PN 	 & 3mm 	 & $<$8.8 	 & \textit{0.3} 	 & -- 	 \\
 	 &  	 & 2mm 	 & $<$10.3 	 & \textit{0.3} 	 & -- 	 \\
 	 & PO 	 & 3mm 	 & $<$4.0 	 & \textit{0.3} 	 & -- 	 \\
 	 &  	 & 2mm 	 & $<$5.4 	 & \textit{0.3} 	 & -- 	 \\
\hline
L1551 	 & PN 	 & 3mm 	 & $<$15.6 	 & \textit{0.9} 	 & -- 	 \\
 	 & PO 	 & 3mm 	 & $<$8.0 	 & \textit{0.9} 	 & -- 	 \\
\hline
Ser-SMM3 	 & PN 	 & 2mm 	 & $<$21.1 	 & \textit{0.9} 	 & -- 	 \\
 	 & PO 	 & 2mm 	 & $<$10.4 	 & \textit{0.9} 	 & -- 	 \\
\hline
Ser-SMM4 	 & PN 	 & 2mm 	 & 50.4 [15.3] 	 & 1.8 	 & 9.4 	 \\
 	 & PO 	 & 2mm 	 & 14.9 [4.8] 	 & 0.6 	 & 7.3 	 \\
\hline
NGC1333-IRAS 4a 	 & PN 	 & 3mm 	 & 37.7 [8.4] 	 & 0.8 	 & 8.7 	 \\
 	 &  	 & 2mm 	 & 103.1 [16.9] 	 & 1.9 	 & 9.3 	 \\
 	 &  	 & 1mm 	 & $<$27.9 	 & \textit{0.8} 	 & -- 	 \\
 	 & PO 	 & 3mm 	 & $<$11.9 	 & \textit{0.8} 	 & -- 	 \\
 	 &  	 & 2mm 	 & $<$13.6 	 & \textit{0.8} 	 & -- 	 \\
 	 &  	 & 1mm 	 & $<$11.1 	 & \textit{0.8} 	 & -- 	 \\
\hline
B1-b 	 & PN 	 & 3mm 	 & 32.9 [7.6] 	 & 0.5 	 & 7.6 	 \\
 	 &  	 & 2mm* 	 & 21.4 [10.7] 	 & 0.5 	 & 7.6 	 \\
 	 &  	 & 1mm 	 & $<$34.0 	 & \textit{1.1} 	 & -- 	 \\
 	 & PO 	 & 3mm 	 & $<$21.4 	 & \textit{1.1} 	 & -- 	 \\
 	 &  	 & 2mm 	 & $<$15.3 	 & \textit{1.1} 	 & -- 	 \\
 	 &  	 & 1mm* 	 & 20.9 [8.4] 	 & 0.5 	 & 10.0 	 \\
\enddata
\tablecomments{Parameters derived from Gaussian fitting of each PO and PN line.  1$\sigma$ uncertainties on the velocity-integrated intensities are listed in brackets.  Tentative detections (2--3$\sigma$) are listed with an *.  Source rest velocities have not been subtracted from the listed velocity offsets.  For non-detections, the integrated intensity upper limit is found from the RMS in a 30 km s$^{-1}$ window centered on the line position (Section \ref{subsubsec:pn_po_columns}).  The line widths used to calculate upper limits are taken to be the average of detected CH$_3$OH lines towards the same source, shown in italics.}
\end{deluxetable}

\begin{figure}[h]
\begin{center}
	{\includegraphics[width=\textwidth]{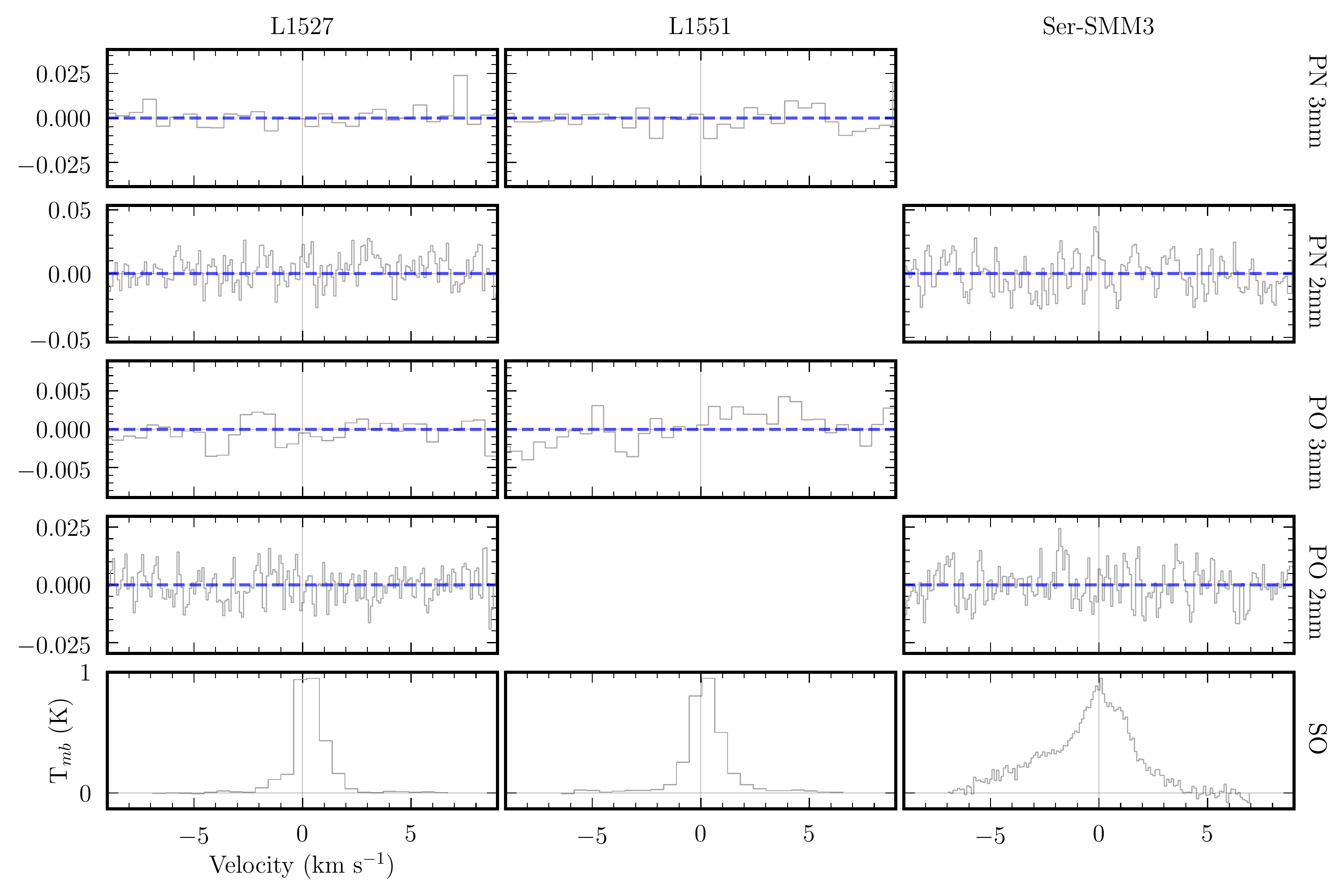}}
    \caption{Non-detections of PN and PO in the sources where no lines were detected. We also include SO lines for comparison: 99.30 GHz for L1527 and L1551, and 138.18 GHz for Ser-SMM3. Ser-SMM3 and L1551 were observed in only one setup each. Velocities are centered on the rest velocity of each source (Table \ref{tab:targets}).}
    \label{fig:nondetections}
\end{center}
\end{figure}

The line fitting results for PN and PO towards all sources are given in Table \ref{tab:p_gaussian_fitting}.  We did not detect phosphorous emission towards three of the seven sources with new observations (L1527, L1551, Ser-SMM3). For these sources, Figure \ref{fig:nondetections} shows each spectrum centered around the expected transitions of PO and PN. For each transition, our fitting routine did not identify any emission line at a significance $>$2$\sigma$. For comparison, we also show SO lines observed towards these sources.

\hspace*{1.5in}

\FloatBarrier
\section{CH$_3$OH and C$^{17}$O Fitting}
\label{app:ch3oh}

\begin{figure}[h]
\begin{center}
	{\includegraphics[width=\textwidth]{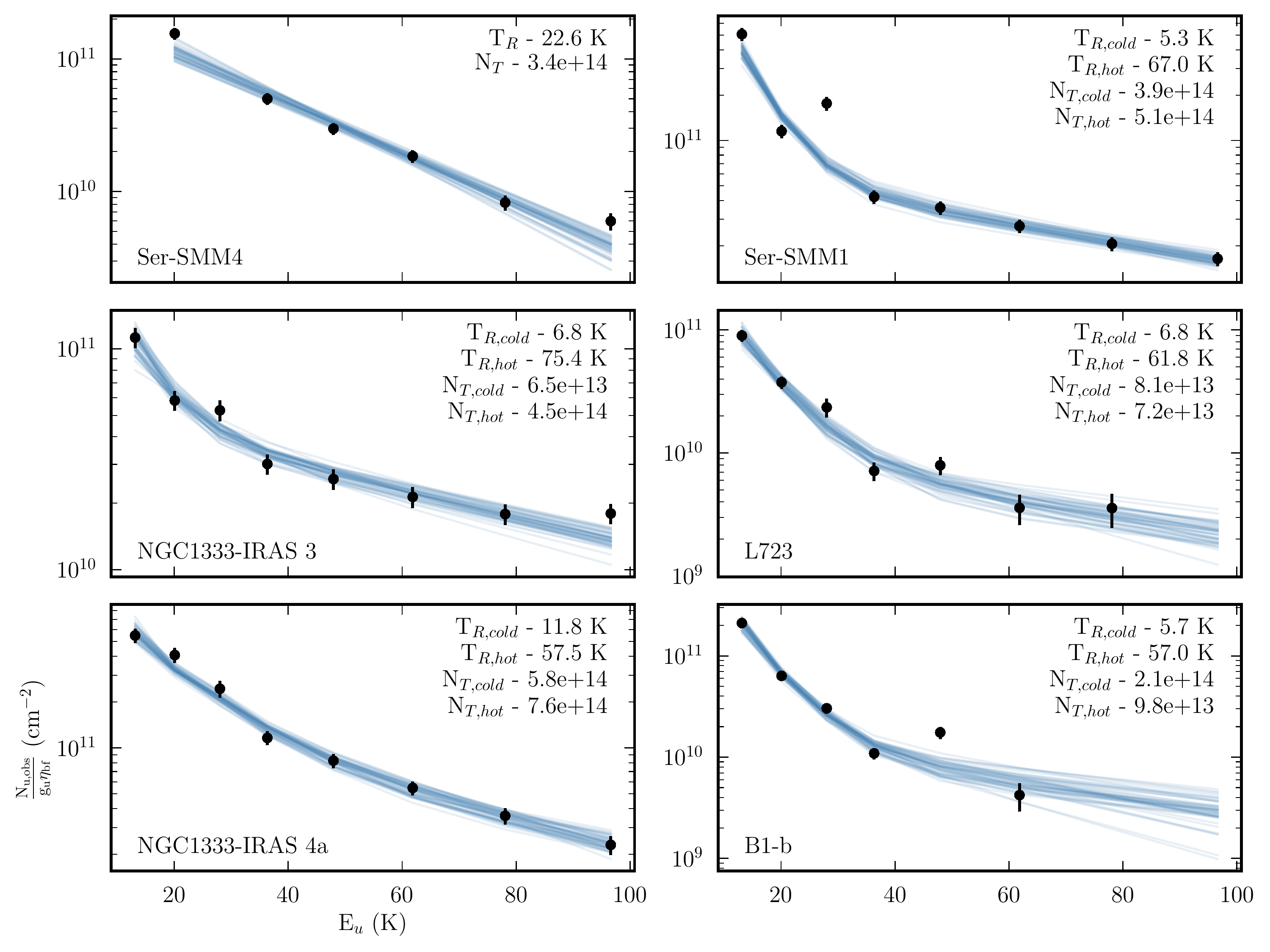}}
    \caption{CH$_3$OH rotational diagrams for the six sources with firm or tentative phosphorus molecule emission. Ser-SMM4 is well fit with a single temperature component, while the other five sources required a two-component fit. The resulting column densities (cm$^{-2}$) and rotational temperatures (K) are listed in the top right corner of each panel, as well as in Table \ref{tab:col_densities}.}
    \label{fig:ch3oh_fitting}
\end{center}
\end{figure}

The results from CH$_3$OH line fitting are listed in Table \ref{tab:CH3OH_detections}.  To find the CH$_3$OH column density in each source, we used the Markov Chain Monte Carlo (MCMC) package \texttt{emcee} \citep{Foreman2013} to fit for both the total column density (N$_T$) and rotational temperature (T$_R$), as described in Section \ref{subsec:columns}.  In three sources (NGC1333-IRAS 3, L723, Ser-SMM1) the CH$_3$OH upper-level populations are not well described by a single rotational temperature. For these sources we fitted the observed upper-level populations as the sum of a hot and cold temperature component with independent column densities. The resulting rotational diagrams for the sources with firm or tentative P molecule detections are shown in Figure \ref{fig:ch3oh_fitting}. 

\begin{figure}[h]
\begin{center}
	{\includegraphics[width=\linewidth]{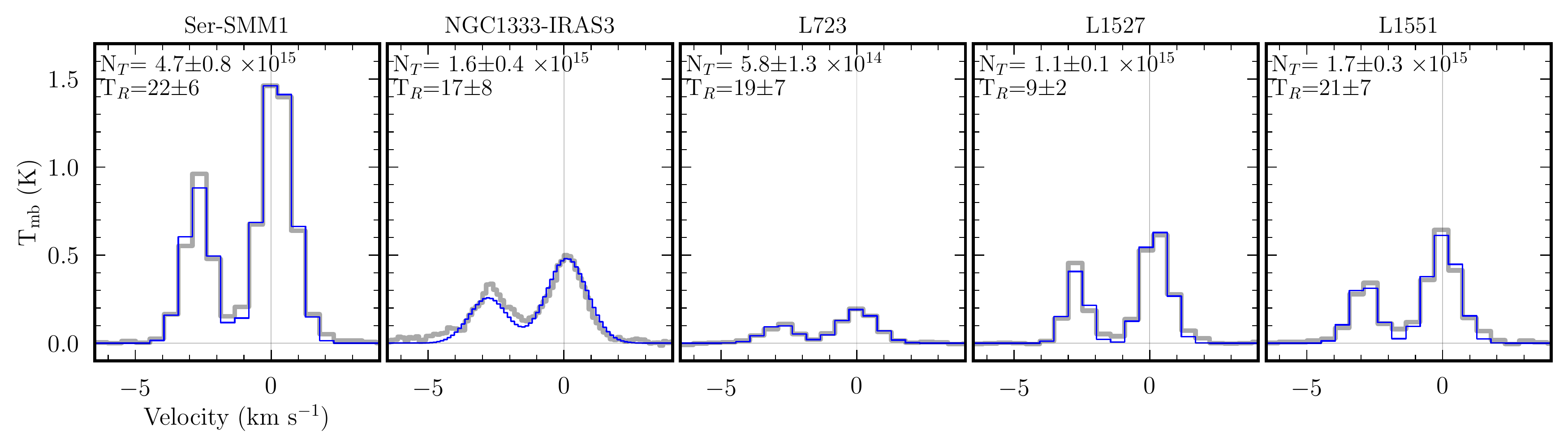}}
    \caption{C$^{17}$O 1--0 observed spectra (grey) and best-fit spectral models (blue).  The resulting column density (cm$^{-2}$) and rotational temperature (K) for each source is listed.}
    \label{fig:c17o_fitting}
\end{center}
\end{figure}

C$^{17}$O 1--0 hyperfine lines were covered for five sources, and were fitted as described in \citet{bergner19} to derive the C$^{17}$O column density and rotational temperature.  The observed spectra and best-fit models are shown in Figure \ref{fig:c17o_fitting}.

\begin{deluxetable}{l|cccc|l|cccc}
\tablenum{8}
\tablecaption{CH$_3$OH Gaussian line fits \label{tab:CH3OH_detections}}
\tabletypesize{\footnotesize}
\tablehead{ \colhead{Source} & \colhead{Frequency} & \colhead{Int.~intensity} & \colhead{Width} & \colhead{Offset} & \colhead{Source} & \colhead{Frequency} & \colhead{Int.~intensity} & \colhead{Width} & \colhead{Offset} \\ 
 & \colhead{(GHz)} & \colhead{(mK km s$^{-1}$)}  & \colhead{(km s$^{-1}$)} & \colhead{(km s$^{-1}$)} & & \colhead{(GHz)} & \colhead{(mK km s$^{-1}$)}  & \colhead{(km s$^{-1}$)} & \colhead{(km s$^{-1}$)}
}
\startdata
L723 & 	 156.829 & 	 84.0 [26.1] 	 & 1.41 	 & 10.6 &  L1551 & 	 96.756 & 	 24.8 [7.6] 	 & 0.57 	 & 6.2 \\
 & 	 157.049 & 	 76.4 [21.2] 	 & 0.86 	 & 10.8 &  & 	 108.894 & 	 175.9 [19.8] 	 & 0.59 	 & 6.4 \\
\cline{6-10}
 & 	 157.179 & 	 148.2 [25.4] 	 & 1.10 	 & 10.8 & Ser-SMM3 & 	 156.489 & 	 68.0 [20.3] 	 & 0.83 	 & 8.1 \\
 & 	 157.246 & 	 112.2 [19.4] 	 & 0.60 	 & 11.0 &  & 	 156.829 & 	 111.8 [25.6] 	 & 1.05 	 & 7.5 \\
 & 	 157.276 & 	 340.9 [39.2] 	 & 0.80 	 & 10.9 &  & 	 157.049 & 	 160.9 [23.9] 	 & 0.69 	 & 7.6 \\
 & 	 96.756 & 	 66.9 [10.4] 	 & 0.88 	 & 10.6 &  & 	 157.179 & 	 272.2 [32.7] 	 & 0.89 	 & 7.4 \\
 & 	 108.894 & 	 230.7 [25.4] 	 & 0.88 	 & 10.6 &  & 	 157.246 & 	 310.8 [36.9] 	 & 0.77 	 & 7.5 \\
 \cline{1-5}
Ser-SMM1 & 	 96.756 & 	 425.7 [43.2] 	 & 0.86 	 & 8.3 &  & 	 157.276 & 	 527.3 [56.1] 	 & 0.78 	 & 7.5 \\
 \cline{6-10}
 & 	 108.894 & 	 1294.7 [129.8] 	 & 0.78 	 & 8.3 & Ser-SMM4 & 	 156.489 & 	 152.1 [22.8] 	 & 1.20 	 & 7.6 \\
 & 	 156.489 & 	 415.0 [44.8] 	 & 1.64 	 & 8.1 &  & 	 156.829 & 	 194.5 [25.4] 	 & 1.44 	 & 7.5 \\
 & 	 156.829 & 	 485.4 [52.3] 	 & 1.64 	 & 8.2 &  & 	 157.049 & 	 393.7 [42.9] 	 & 1.31 	 & 7.6 \\
 & 	 157.049 & 	 575.5 [60.1] 	 & 1.38 	 & 8.2 &  & 	 157.179 & 	 556.7 [58.1] 	 & 1.29 	 & 7.5 \\
 & 	 157.179 & 	 666.9 [68.3] 	 & 1.41 	 & 8.3 &  & 	 157.246 & 	 785.0 [80.2] 	 & 1.19 	 & 7.4 \\
 & 	 157.246 & 	 664.7 [68.7] 	 & 1.56 	 & 8.5 &  & 	 157.276 & 	 1406.1 [141.9] 	 & 1.13 	 & 7.5 \\
  \cline{6-10}
 & 	 157.276 & 	 1050.8 [106.2] 	 & 1.10 	 & 8.4 & NGC1333- & 	 96.756 & 	 706.1 [87.9] &	0.98 	 & 7.8 \\
 \cline{1-5}
NGC1333- & 	 96.756 & 	 151.9 [16.8] 	 & 1.47 	 & 8.1 & IRAS 4a & 	 108.894 & 	 1396.9 [147.9] 	 & 0.80 	 & 8.0 \\
 IRAS 3& 	 108.894 & 	 286.5 [30.1] 	 & 1.09 	 & 8.2 &  & 	 156.489 & 	 586.9 [82.5] 	 & 1.33 	 & 8.0 \\
 & 	 156.489 & 	 457.4 [48.7] 	 & 1.99 	 & 8.0 &  & 	 156.829 & 	 844.3 [104.4] 	 & 1.19 	 & 7.7 \\
 & 	 156.829 & 	 421.3 [45.4] 	 & 1.86 	 & 8.0 &  & 	 157.049 & 	 1163.2 [123.7] 	 & 0.98 	 & 7.7 \\
 & 	 157.049 & 	 455.3 [50.5] 	 & 1.87 	 & 8.0 &  & 	 157.179 & 	 1539.7 [164.6] 	 & 0.83 	 & 7.8 \\
 & 	 157.179 & 	 480.9 [51.3] 	 & 1.75 	 & 8.2 &  & 	 157.246 & 	 1829.1 [192.3] 	 & 0.77 	 & 7.7 \\
 & 	 157.246 & 	 473.0 [50.3] 	 & 1.54 	 & 8.2 &  & 	 157.276 & 	 3696.5 [417.4] 	 & 0.54 	 & 7.6 \\
  \cline{6-10}
 & 	 157.276 & 	 532.7 [54.9] 	 & 1.06 	 & 8.2 &  L1527 & 	 108.894 & 	 101.1 [14.7] 	 & 0.36 	 & 5.8 \\
\cline{1-5}
B1-b & 	 96.756 & 	 87.6 [10.2] 	 & 0.53 	 & 7.5 &  & 	 157.276 & 	 96.0 [13.6] 	 & 0.23 	 & 5.9 \\
 \cline{6-10}
 & 	 108.894 & 	 538.5 [62.7] 	 & 0.55 	 & 7.4 \\
 & 	 157.049 & 	 89.9 [28.3] 	 & 0.97 	 & 8.5 \\
 & 	 157.179 & 	 327.6 [45.5] 	 & 1.45 	 & 5.7 \\
 & 	 157.246 & 	 171.3 [22.1] 	 & 0.48 	 & 7.2 \\
 & 	 157.276 & 	 576.3 [59.6] 	 & 0.39 	 & 7.2 \\
\enddata
\tablecomments{Parameters derived from Gaussian fitting of CH$_3$OH lines.  1$\sigma$ uncertainties on the velocity-integrated intensities are listed in brackets.  Source rest velocities have not been subtracted from the listed velocity offsets.  }
\end{deluxetable}

\FloatBarrier
\newpage
\bibliography{references}{}
\bibliographystyle{aasjournal}

\end{document}